\documentclass[10pt,journal,compsoc]{IEEEtran}
\usepackage{url}
\usepackage{caption}
\usepackage[noadjust]{cite}
\usepackage{graphicx}
\usepackage{multirow}
\usepackage{comment}
\usepackage{makecell}
\usepackage{ragged2e}
\usepackage{rotating}
\usepackage{lscape}
\usepackage[table]{xcolor} 
\usepackage{nomencl}
\usepackage{dblfloatfix}
\makenomenclature
\usepackage{wrapfig}
\usepackage{makecell}
\usepackage{balance}
\usepackage[utf8]{inputenc}
\usepackage{ragged2e}
\usepackage{booktabs}

\usepackage{pifont}
\usepackage{tabulary}
\usepackage{comment}
\usepackage{lipsum}
\usepackage{array, tabularx, multirow, booktabs, diagbox}
\usepackage{adjustbox}

\ifCLASSOPTIONcompsoc
\else
\fi
\ifCLASSINFOpdf
\else
\fi

\begin{document}
	\title{Energy Harvesting in 5G Networks: Taxonomy, Requirements, Challenges, and Future Directions}
	\author{Muhammad~Imran,~
		Latif U. Khan,~
		Ibrar~Yaqoob,~\IEEEmembership{Senior Member,~IEEE,}
		Ejaz~Ahmed,~\IEEEmembership{Senior Member,~IEEE,}
		Muhammad~Ahsan~Qureshi,~and
		Arif~Ahmed

		\IEEEcompsocitemizethanks{\IEEEcompsocthanksitem Muhammad Imran is with the College of Applied Computer Science, King Saud University, Riyadh, Saudi Arabia.
			
			\IEEEcompsocthanksitem Latif U. Khan is  with the Department of Telecommunication Engineering, University of Engineering \& Technology, Mardan, Pakistan.
			
			\IEEEcompsocthanksitem Ibrar Yaqoob is with the Department of Computer Science and Engineering, Kyung Hee University, Yongin-si 17104, South Korea.
			
			\IEEEcompsocthanksitem Ejaz Ahmed is with the Centre for Mobile cloud Computing Research (C4MCCR), University of Malaya, 50603 Kuala Lumpur, Malaysia. 
			
			\IEEEcompsocthanksitem Muhammad Ahsan Qureshi is with Faculty of Computing and Information technology, University of Jeddah, Khulais, Saudi Arabia. 
			
			\IEEEcompsocthanksitem Arif Ahmed is with Department of Computer Science \& Engineering, National Institute of Technology, Silchar, Silchar, India.

		}
\thanks{
}}
\IEEEcompsoctitleabstractindextext{
\justify	
		\begin{abstract}
			Consciousness of energy saving is increasing in fifth-generation (5G) wireless networks due to the high energy consumption issue. Energy harvesting technology is a possible appealing solution for ultimately prolonging the lifetime of devices and networks. Although considerable research efforts have been conducted in the context of using energy harvesting technology in 5G wireless networks, these efforts are in their infancy, and a tutorial on this topic is still lacking. This study aims to discuss the beneficial role of energy harvesting technology in 5G networks. We categorize and classify the literature available on energy harvesting in 5G networks by devising a taxonomy based on energy sources; energy harvesting devices, phases, and models; energy conversion methods, and energy propagation medium. The key requirements for enabling energy harvesting in 5G networks are also outlined. Several core research challenges that remain to be addressed are discussed. Furthermore, future research directions are provided.
		\end{abstract}
		
		}
	\maketitle
	\IEEEdisplaynotcompsoctitleabstractindextext
	\IEEEpeerreviewmaketitle

	\section{Introduction}
	With the fruition of Internet of Things (IoT) and machine-to-machine communication, the design and implementation of emerging fifth-generation (5G) networks must be tailored accordingly. These 5G networks are projected to be widely deployed in 2020 to enable massive connectivity and provide data rate speeds of 10 Gbps in the peak hours for low mobility and 1 Gbps of data rate for high mobility \cite{ercan2018rf}. With the significant reduction of delay, support for real-time multimedia applications is likely to increase but satisfying stringent energy and computation constraints in an affordable and sustainable way is a real challenge \cite{mekikis2018connectivity}. To cope with these challenges, 5G networks are expected to deploy low-powered radio access nodes, which form small cells across a region. These radio access nodes can work in licensed and unlicensed spectrum bands, thereby increasing energy consumption in the network infrastructure \cite{shah20185g}.\par
	The contribution of CO2 emission by the ICT in 2015 was 5\% and expected to increase due to proliferation of a sheer number of mobile devices in upcoming years. By 2020, 75\% of the ICT sector would be wireless, thereby indicating that wireless communications will be the critical sector for researchers as far as reducing ICT-related CO2 emissions is concerned. This condition will trigger the innovations in the network architecture and technologies of wireless communication, thereby leading to 5G cellular networks \cite{wang2018new}. Thus, providing and supplying energy from other sources to support the high rates and continuous availability of the network and designing energy-efficient architectures are essential.\par
\begin{figure*}[!t]
		\centering
		\captionsetup{justification=centering}
		\includegraphics[width=12cm, height=6cm]{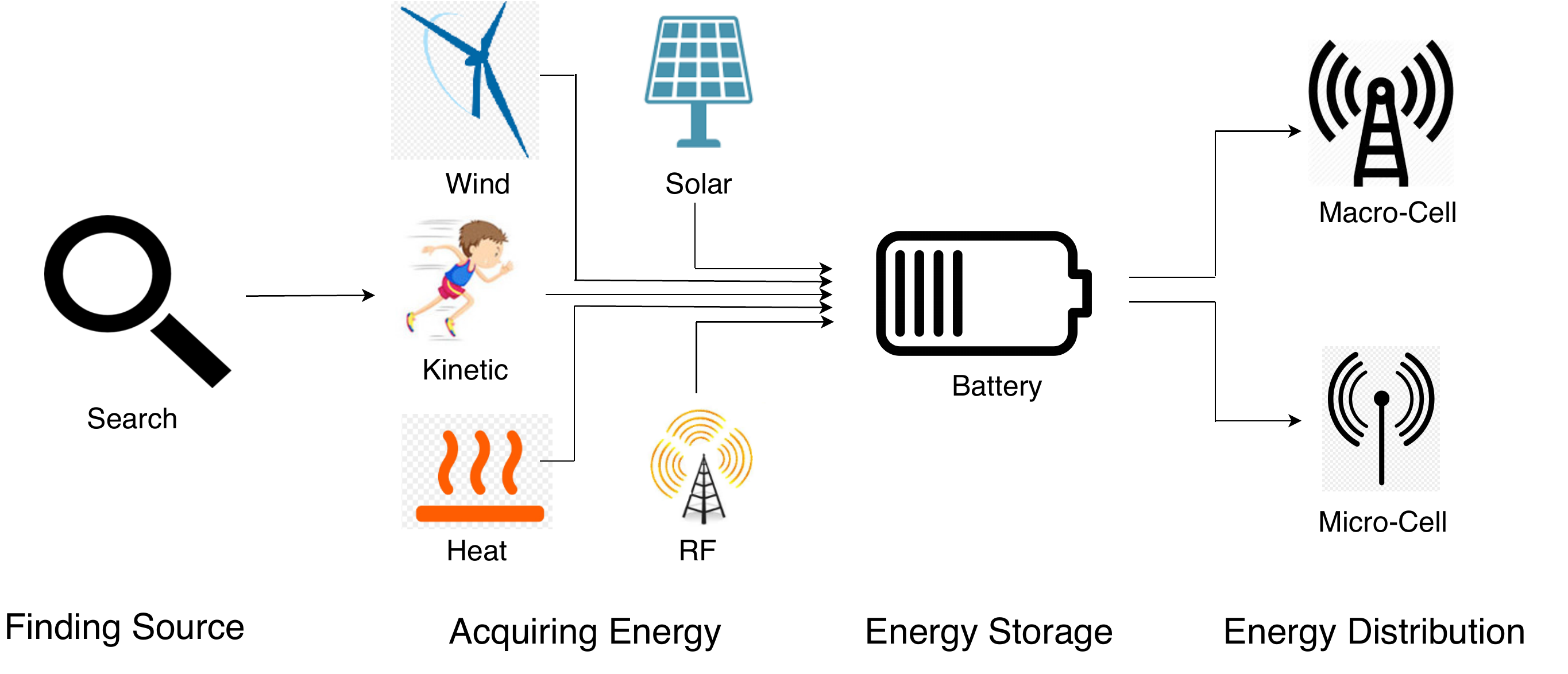}
		\caption{Energy harvesting process in 5G networks.}
		\label{fig:Energy harvesting process}
	\end{figure*}

	Energy harvesting techniques can be used to produce energy from the surrounding environment, thereby converting energy to electrical power for use in 5G network devices, such as base stations (BSs) and mobile phones \cite{sinaie2018delay}. Figure~\ref{fig:Energy harvesting process} shows the process of energy harvesting in 5G networks. Energy harvesting is a promising technology that does not diminish energy consumption of devices but enables a device to be self-powered when emergency power shortage is encountered.\par  
	Although several studies have been conducted on 5G networks in terms of energy harvesting, a tutorial on this topic is lacking. The contributions of this study are as follows:\par
	\begin{itemize}
		\item We present motivation for employing energy harvesting in 5G networks.
		\item We provide an overview of several possible ways of harvesting energy in 5G multi-tier heterogeneous network architecture.  
		\item We taxonomize the literature based on indispensable parameters.	
		\item We outline the key requirements of deploying energy harvesting solutions in 5G networks.
		\item Several open challenges and future directions in enabling energy harvesting on 5G networks are identified, enumerated and discussed.
		
	\end{itemize}
	These contributions are provided in separate sections from 2-7. We provide concluding remarks in section 8.
	\section{Motivation}
	\label{sec:motivation}
	Various applications, ranging from basic communication and social networking to infotainment services that utilize mobile network for their effective usage, are currently available. These applications demand high  data transfer rate and massive connected network. The ever-increasing demand of faster mobile communication and utilization of increasing data paved the way toward 5G mobile networks. The 5G network provides higher data transfer rate and improved coverage than its predecessors. However, the quest toward faster and more reliable mobile communication has resulted in the overall energy consumption stipulation in 5G networks \cite{guntupalli2018demand}. One the basis of the extrapolation of important metrics, such as number of communication devices sold per year and data traffic, communication technology could use 51\% of the global electricity by 2030 if adequate improvement in electric efficiency is unachieved. \par
	Typical household and office environments are considerably changed in terms of energy requirements, especially in the past decade as increasing wireless nodes are being added to the network. A total of 4.43 billion mobile phone subscribers exist worldwide as opposed to 4.01 and 4.23 billion in 2013 and 2014, respectively, thereby indicating an approximately 5\% annual increase. The external and internal networks with a sufficient quantity of battery-powered wireless nodes require energy to be harvested \cite{chen2017real}.\par
	The development of energy-efficient algorithms, standards, and protocols is a passable way that decreases energy consumption in mobile communication \cite{nguyen2016secure}. However, one of the most adequate measures to fulfill the increasing energy requirements is to use renewable energy sources. Researchers have explored numerous ways to obtain adequate power for wireless networks.  Energy harvesting involves capturing and storing energy derived from external sources, including solar, thermal, wind, vibration, and human body energies, for wireless devices. These sources are commonly and abundantly available in the environment; hence, energy harvesting is the most promising technique to fulfill the energy requirements of wireless networks.\par
	\section{5G Multi-tier Heterogeneous Network Architecture}
	\label{sec:archi}
\begin{figure}[!h]
		\centering
		\includegraphics[width=8cm]{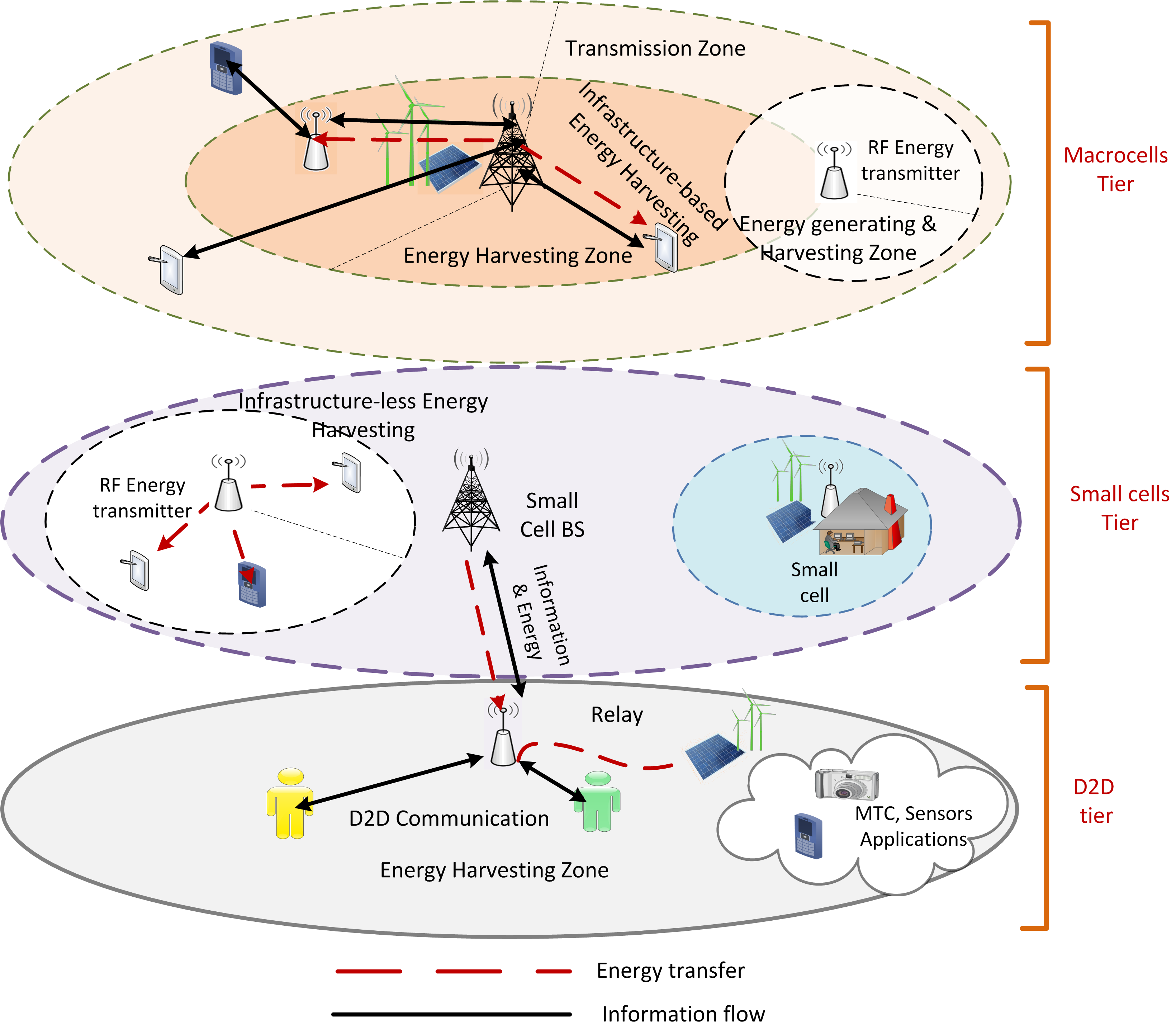}
		\caption{5G HetNets with energy harvesting.}
		\label{fig:5GHetNets}
	\end{figure}

	The 5G cellular network is envisioned to enable users with novel applications, such as augmented reality, self-driving cars, smart homes, smart farming, and smart e-health care. All 5G applications are generally divided into three use cases identified by the International Telecommunication Union Radio communication Sector. These use cases include enhanced mobile broadband (eMBB), ultra-reliable low-latency communication (URLLC), and massive machine-type communication (mMTC). In these use cases, mMTC is characterized by a massive number of nodes with low sensitivity to latency. The URLLC has the lowest latency requirements among the three uses cases with enhanced reliability and includes various applications, such as self-driving cars and mission critical applications. Meanwhile, eMBB has high throughput requirement applications, such as HD  video streaming, virtual, and augmented reality. To enable these use cases, 5G networks will use heterogeneous networks (HetNets) along with densification of small-cell BSs and device-to-device (D2D) communication. The results of densification are high spectral efficiency and decreased mobile transmission power. However, these results will cause an increase in the overall energy consumption of networks. Therefore, fulfilling the energy demands of 5G networks is imperative, as shown in Figure~\ref{fig:5GHetNets}.\par 
	The different tiers of 5G HetNets can be empowered by  grid station and energy harvesting. Using harvested energy and grid power simultaneously is highly effective. Power from energy harvesting and grid is used because of the  substantial random variations that exist in harvested energy. Therefore, grid energy, in addition to harvested energy, is necessary to fulfill the energy requirements of the different tiers of 5G networks.\par
	Macrocell BSs can be used to transmit information and energy simultaneously. This technique enables mobile devices with limited energy to fulfill their energy requirements. Radio frequency (RF) energy harvesting zone is smaller than the transmission zone because the former requires higher energy than the that required for information detection. In addition to RF energy harvesting from macrocell BSs, RF energy harvesting sources with a low range can be used to supply energy to devices in its proximity. Interference signals can also be used for energy harvesting, which improves energy efficiency of cellular networks in contrast to system performance degradation due to interference.\par
	The D2D tier in 5G cellular networks can be used to improve the system throughput. Harvested energy along with devices that have battery sources can be leveraged to enable energy-efficient D2D communication. Furthermore, D2D communication can be assisted by relays to enhance further their performance, especially during poor channel conditions. The third-layer relays can be connected to the macrocell eNB  and then used to serve D2D and mobile users with poor channel conditions \cite{5Ghetnets}. The relays can also be operated by harvesting energy from the environment in addition to RF energy and then simultaneously assist D2D communication and communication between the user and the BS.\par
    \section{Taxonomy of 5G Networks with Energy Harvesting}
	\label{sec:taxonomy}
\begin{figure*}[!t]
		\centering
		\captionsetup{justification=centering}
		\includegraphics[width=14cm, height=9cm]{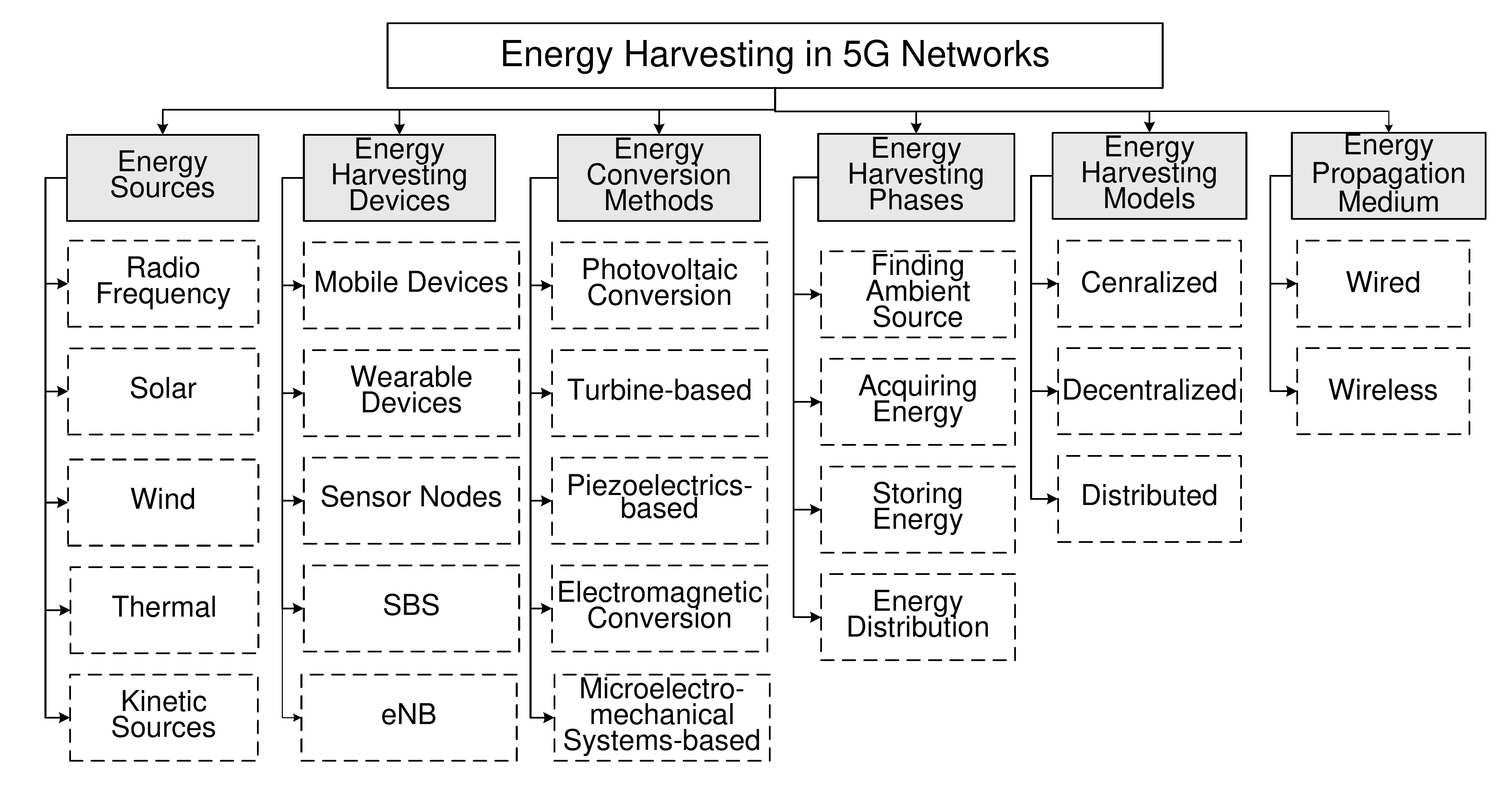}
		\caption{Taxonomy of energy harvesting in 5G networks.}
		\label{fig:TaxEnergyHarvesting}
    	\end{figure*}

	The taxonomy on energy harvesting in 5G network is shown in Figure \ref{fig:TaxEnergyHarvesting}. This taxonomy is categorized on the basis of the following attributes: a) harvesting technologies, b) harvesting devices, c) energy conversion methods, d) harvesting phases, e) energy harvesting models, and f) energy propagation medium.\par
	\subsection{Energy Sources} 
	Many energy sources are available in the surroundings for energy harvesting in 5G networks \cite{lee2018multi}. These energy sources can be classified into five main categories: a) RF, b) solar, c) wind, d) thermal, and e) body temperature.\par
	The RF-based harvesting uses RF signals in the air to produce energy. This type of energy is mined from the electromagnetic radiations generated by different sources, such as telecom BS, Bluetooth, and infrared devices. Solar energy is an established technology for energy generation at a large scale. Photovoltaic systems are used to produce electricity from solar energy. Electricity can be produced from milliwatt to megawatt range. In addition, energy can be produced from wind using wind turbines that convert kinetic energy of the wind into mechanical power. Thereafter, the mechanical power is transformed into electricity by using generators. The electricity through wind energy can be produced from kilowatt to megawatt range. The energy can be rapidly produced from various thermal sources, such as animals, persons, and machines, using a thermoelectric generator. This power can be classified as thermal energy. Energy harvested from kinetic sources can be a key enabling mechanism for emerging 5G-based IoT and wearable devices. A kinetic energy harvester converts the mechanical power obtained from the movement of an object into electrical energy using a micro-electromechanical system. \par	
	\subsection{Energy Harvesting Devices}
	The harvesting devices in 5G networks range from the user mobile devices connected to the 5G network to the telecom network operator’s BSs. The mobile devices are one of energy-drained devices with limited power storage capacity  \cite{moradian2018tradeoff}. However, mobile devices can leverage the vibration and thermal harvesting mechanisms to produce energy from the human body. Similarly, wearable devices can leverage body temperature and movement to mitigate the energy scarcity problem. Although SBS  and eNB are supposed to deploy densely in 5G networks, they will require substantial energy for power. These deployments are in underdeveloped regions, such as hill terrain and frequently flood-affected areas. In such places, BSs of 5G network can utilize the energy harvesting technologies to acquire wind, solar, and thermal energies. \par
	\subsection{Energy Conversion Methods} 
	The performance of energy harvesting in 5G is considerably dependent on the available energy to harvest and energy conversion technique used (i.e., the efficiency of energy conversion from the source ambient to the usable electric energy). Energy conversion techniques are related with the power sources. Some key energy conversion techniques are photovoltaic and electromagnetic conversion and turbine-, piezoelectric-, and microelectromechanical system-based. Photovoltaic conversion is used to transform light into electricity by semiconducting materials that possess photovoltaic characteristics. A turbine is a rotary turbomachine that transforms kinetic energy into electrical power. An example of a turbomachine is wind turbine. A piezoelectric-based technique collects the electric charge in a solid material resulting from a mechanical stress. Electromagnetic energy conversion is an important mechanism for energy harvesting. The conversion requires an energy conversion device and adapter. Microelectromechanical systems are small devices comprised of integrated electrical and mechanical components that are used to produce a small amount of energy by vibration and movement of an object.\par	
	\subsection{Energy Harvesting Phases} 
	The energy harvesting process in 5G can be divided into four phases: a) finding ambient source, b) acquiring energy, c) energy storage, and d) energy distribution phase. In the ambient source finding phase, the harvesting device scans the surrounding environment to find any harvesting energy sources. The harvesting devices will switch on the conversion device to acquire energy on the basis of the available ambient sources. The next phase is transferring the energy from the conversion device to the storage section to store the energy for future utilization. The last phase is to distribute the acquired energy to the final harvesting devices.\par	
	\subsection{Energy Harvesting Models} 
	The solutions of energy harvesting can be centralized, decentralized, and distributed. In the centralized energy harvesting model, a central harvesting device is responsible to harvest the energy in the region (nearby BSs), and the 5G devices and energy storage components are also co-located with each other. In this type of solution, the devices remain in static, and the environment is easily predictable, thereby indicating that the complexity is low. The central devices dissipate the energy to the surrounding environment. However, in the decentralized energy harvesting solution, the devices are in a random state. The devices individually harvest the energy, and no or less communication exists among the devices (such as sensor nodes and mobile devices) in the region. Finally, distributed energy harvesting is similar to the centralized harvesting solution; however, the number of harvesting devices is more than one. Harvesting devices coordinate with each other.\par
	\subsection{Energy Propagation Medium} 
	Similar to information flow in the 5G network, the harvested energy can be distributed among the peer devices in wired and wireless medium. RF energy is transferred around the environment in a radio wave form. The harvesting device in this category, such as wireless mobile charger, uses a wireless energy propagation medium. In the wired medium, the energy is transmitted in the physical wire from one point to another (e.g., from conversion device to the battery storage). \par
    \section{Energy Harvesting Requirements for 5G Networks}
	\label{sec:requirements} 
\begin{figure*}[!t]
		\centering
		\captionsetup{justification=centering}
		\includegraphics[width=14cm]{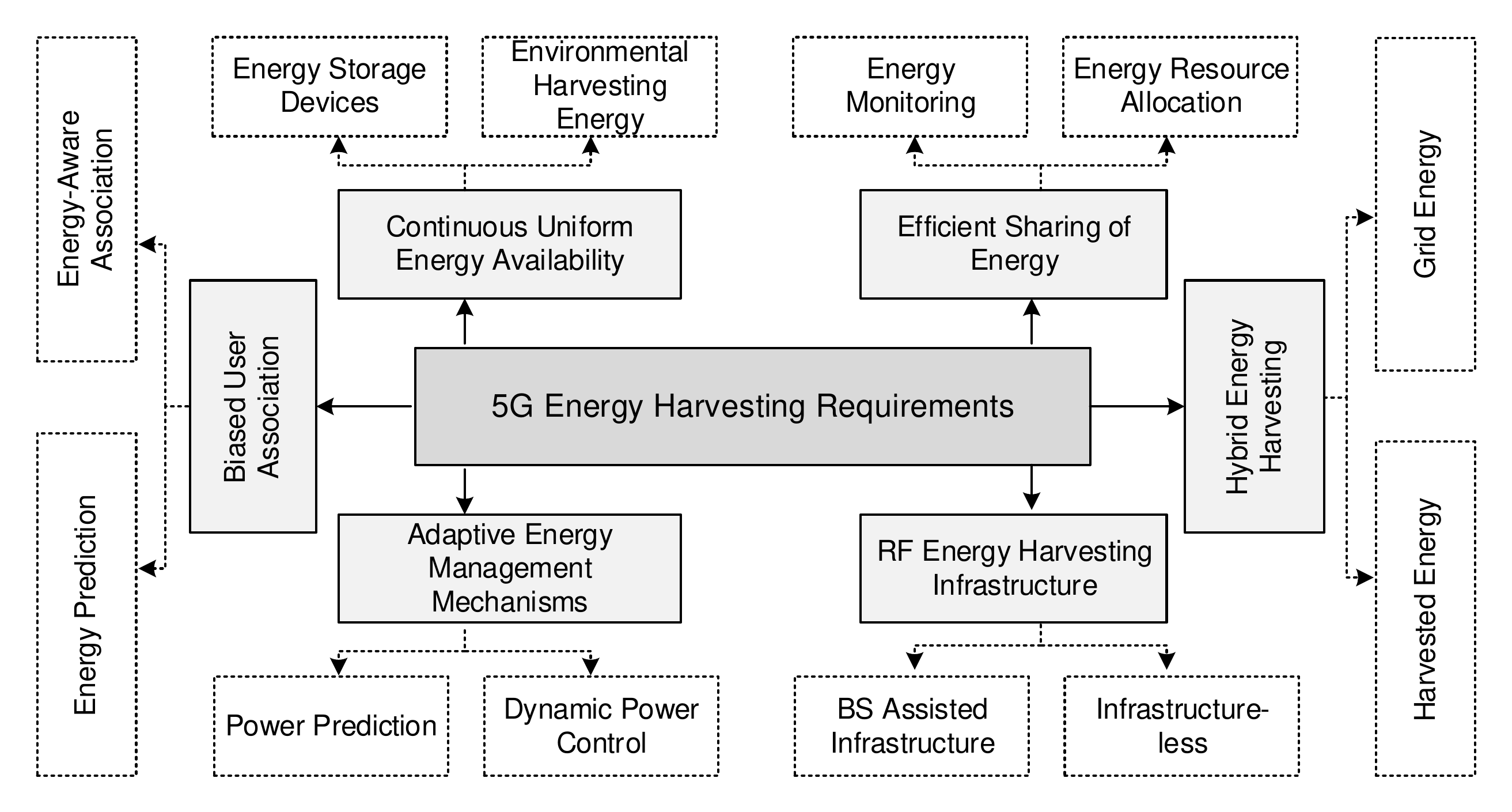}
		\caption{Energy harvesting requirements for 5G networks.}
		\label{fig:requirements}
	\end{figure*}

	Figure~\ref{fig:requirements} summarizes the key requirements for enabling energy harvesting in 5G cellular networks. Further details are provided in the following subsections. \par
	\subsection {Continuous Uniform Energy Availability} 
	Energy harvesting produce power supply without any interruption and produce energy from the surroundings without using any power supply from other sources. The scarcity of energy harvesting will be prevalent in 5G networks and in cases of emergency situation, such as flood and earthquake, electric supply is unavailable to power 5G network equipment from nowhere. In such a situation, energy drawn from the surrounding is the only option left to power 5G devices. Given the variations in harvesting energy, we must use energy storing devices to enable the continuous energy supply to 5G network devices. For example, if harvested energy is the only source in a disaster situation and we are using solar energy harvesting, then we will face the challenge of variations in the harvested energy and run out of energy at night time or during a cloudy weather. To cope with above discussed potential issues, the harvesting energy device must be equipped with a battery to enable its continuous uniform operation.\par
	\subsection{Efficient Sharing of Energy} The 5G network can use the harvested energy and energy derived from the main power supply, and they can be transmitted in wired line and wireless connection, such as a wireless charger. A new efficient energy sharing protocol needs to incorporate several 5G network performance parameters, such as bandwidth and delay, to share the harvested energy because energy harvesting is dependent on the availability of resources in the environment. Various schemes based on optimization theory and game theory can be used to ensure efficient sharing of harvested energy. On the other hand, a hybrid scheme using both optimization theory and game theory can be leveraged to enable efficient sharing of harvested energy. \par
	\subsection{Adaptive Energy Management Mechanisms} Generally, energy harvesting system in 5G network involves power harvesting devices, battery for energy storage, and supercapacitors. The system can monitor the electric consumption level of the various components and the power level of the battery with energy harvesting devices. On the other hand, both environmental and harvested energy have limitations in their availability. Therefore, we must develop an adaptive energy management mechanism that will provide dynamic power from the harvested energy depending on the energy requirement. The two main aspects of adaptive energy mechanism are energy consumption prediction of the operating device and dynamic power control of the harvesting device.\par
	\subsection{RF Energy harvesting Infrastructure}
	In 5G cellular networks, presence of wide-band communication signals offer opportunity to harvest RF energy. The 5G user devices generally have energy limitations compared to BS and access point (AP). Therefore, AP and BS can be utilized to jointly transmit energy and information signals. The RF energy harvesting architecture can be either infrastructure-based or infrastructure-less. In the infrastructure-based architecture, the centralized BS or AP transmits RF energy and information. The users communicate one another through the same centralized BS or AP from which they harvest energy. By contrast, infrastructure-less architecture involves the utilization of wireless energy transmitters, and users in its vicinity can use this RF energy. The users can communicate with one another while using energy from the RF energy source to continue its operation in a cost-effective way. New protocols should be designed to enable the architecture of 5G that supports RF energy harvesting in infrastructure-based and -less modes. \par        
	\subsection{Hybrid Energy harvesting}
	5G devices empowered by harvesting energy sources must be able to withstand the device power requirements. However, there exist significant variations in RF energy. Additionally, environmental harvesting energy has both variations and outage issue. For instance, harvesting energy from the sun is possible only at day times. To cope with variations in harvesting energy sources, it is necessary to use hybrid energy sources that jointly utilize the harvested energy and energy from the power grid and diesel generators. Such a hybrid energy harvesting system should fully utilize the harvested energy. In addition, the system should use the energy from power if required for the operation of 5G devices.  \par
	\subsection{Biased User Association}
	In 5G networks, HetNets are used to meet the high data rate and latency requirements. HetNets consist of multiple tiers, such as macrocell and child tiers. The transmission power of the child tier is lower than the parent one. Therefore, the HetNets empowered by energy harvesting suffer from the challenge of unbalanced user association because the user association based on the received signal strength might result in overloading of the macrocell tier. This type of unbalanced operation in cellular networks empowered jointly by energy harvesting and other sources results in inefficient energy consumption. To avoid such type of situation, it is necessary to consider biased user association in 5G cellular networks. A biased user association that achieves a load-balanced operation should be performed to prevent overloading of the macrocell tier. Alternatively, we can say that it results in energy-aware user association.\par
	\section{Open Research Challenges}
	\label{sec:openchallenges}
	The following discussion highlights the challenges involve in enabling energy harvesting on 5G networks that must be overcome. Table 1 summarizes the research challenges along with their perceived solutions.\par
	\begin{table*}
		\caption {Summary of the research challenges and their perceived solutions.} \label{tab:challenges} 
		\centering
		\resizebox{\textwidth}{!}{
			\begin{tabular}{p{4cm}p{6cm}p{6cm}}
				\toprule
				Challenges & Causes & Guidelines \\
				\toprule
				\textbf{Broadband Energy Harvesting} & \begin{itemize} \item Wide range of spectrum usage in 5G  \item High interference at cell edges \end{itemize} & \begin{itemize} \item Energy harvesters with high range of operating frequencies   \end{itemize} \\
				\midrule 
				
				\textbf{RF Energy Harvesting Relays} & \begin{itemize} \item Existence of relays for throughput enhancement \item Significant variations in environmental harvested energy \end{itemize}& \begin{itemize} \item Novel joint communication and energy harvesting protocols for relays \item Optimal placement of relays based on joint optimization of energy harvesting and communication  \end{itemize}\\
				\midrule
				\textbf{Online Energy Harvesting} & \begin{itemize} \item Harvested energy outage \item Frequent variations in both RF and environmental harvesting energy sources  \end{itemize}& \begin{itemize} \item Stochastic optimization-based techniques  \item Harvesting schemes based on machine learning  \end{itemize}\\
				\midrule   
				\textbf{Interference-Assisted Energy Harvesting} & \begin {itemize} \item Presence of interference signals in wireless medium from a variety of sources \item Environmental harvested energy limitations \end{itemize}& \begin{itemize} \item Novel transceiver design for joint communication and energy harvesting    \end{itemize}
			
			\\
			\midrule
			
			\textbf{Harvested Energy Resource Allocation} & \begin{itemize} \item Limited harvested energy \item High latency in harvested energy resource allocation \end{itemize} & \begin{itemize} \item Adaptive harvested energy resource allocation \item Optimization theory based harvested energy resource allocation \item Game theory based harvested energy resource allocation \item Learning theory based harvested energy resource allocation\end{itemize} \\ 
			
			\bottomrule
		\end{tabular}
	}
\end{table*}
\subsection{Broadband Energy Harvesting}
In a 5G cellular network, novel frequency bands in addition to predecessor cellular networks bands are expected to use for enabling superfast access. Numerous devices operating at a wide range frequency bands offer opportunity for other devices to harvest their RF energy. More specifically, the mobile devices that are on the edge of a cell may experience interference from the neighboring cells. The edge devices need more power to obtain a sufficient level of signal-to-noise ratio while transmitting on the uplink. This energy requirement can be fulfilled by harvesting the energy from ambient RF neighboring cell sources. However, this approach needs a new receiver architecture with circuits to harvest the energy from signals of wide range of frequencies. The unreliable and dynamic nature of the RF ambient sources make it challenging to harvest and store the sufficient amount of energy for meeting the edge device requirement.\par
\subsection{RF Energy Harvesting Relays}
5G cellular networks are intended to use relay for throughput, especially at the cell edges. These relays have energy constraints and must be powered by harvested energy. Both energy from environmental sources and RF sources can be used. However, there exist significant variations in the environmental sources, such as sun-light and wind energy. Therefore, RF energy harvesting relays is a feasible solution to enable simultaneous communication and RF energy harvesting. In cellular networks, the intermediate relay node can use the source signal energy to forward it to the destination. Determining the positions and selection of relays for improving overall performance is an open research area for 5G cellular networks given that energy harvesting is used in cellular networks for achieving an energy-optimized operation. Other than optimal placement of relays, novel protocols offering joint communication and energy harvesting performance enhancement must be proposed. \par
\subsection{Online Energy Harvesting}
Energy in wireless communication can be harvested either from the environment, such as wind and sun, or RF signals. In both cases, the variations that exist in the available energy of the harvesting sources (e.g., wind, sun, and RF signals) will impose limitations on the device design that range from a user equipment to the operator BS. The device energy requirement at a certain time should not exceed the harvested energy up to that time. Energy harvesting techniques can work in an offline fashion using prior assumption of the known harvested energy from the source. However, obtaining the information about harvested energy in practical scenarios is difficult. Therefore, efforts must be made to design the effective online approaches for 5G, which are not based on the prior assumption of the amount of energy harvested, to overcome the limitation of the offline approach. Stochastic optimization-based techniques that utilize the assumption of the known energy process statistics can be used. In addition to stochastic optimization, techniques based on learning theory can also be used for online energy harvesting.\par
\subsection{Interference-Assisted Energy Harvesting}
In 5G cellular networks, the existence of a variety of interference signal offers the opportunity to harvest their RF energy. Additionally, environmental energy harvesting has limitations that impose challenges on devices operating via harvested energy only. Although interference has degraded the performance of the communication systems, it can be used positively in energy harvesting. Wireless signal contains energy and information; therefore, we can harvest energy from the interference signal. In 5G HetNets, different tiers exist, such as macrocell and small-cell tiers, in which interference is a prominent issue, especially at the macrocell edges. On the other hand, there exist significant limitations in the environmental energy harvesting. Therefore, we can use interference in energy harvesting to improve the overall system performance. The architecture of energy harvesting devices that range from user devices to network operator BSs should be designed in a way that it effectively utilizes interference in energy harvesting. \par
\subsection{Harvested Energy Resource Allocation}
Harvested energy from both environmental and RF sources has significant limitations. Therefore, they must be used efficiently to fulfill need of 5G networks. The manner in which energy resources from different energy harvesting sources should be allocated among 5G devices to maximize jointly the overall profit and QoS must be determined. Numerous harvested energy resource allocation schemes suffer from high latency due to their associated computational complexity. To address this concern, designing algorithms that will perform efficient resource allocation with low latency to maximize the overall quality of experience is necessary to address the aforementioned concern. Energy harvesting systems can obtain power either from natural sources or RF signals. Therefore, we must design adaptive algorithms to allocate the energy resources among the 5G network devices. Several schemes based on game theory, learning theory, and optimization theory can be used for efficient harvested energy resource allocation. On the other hand, an adaptive harvesting energy resource allocation scheme based jointly on game theory and learning can be a viable solution.\par
\section{Future Directions}
This section provides several indispensable future directions to new researchers working in the domain.\par
\subsection{Energy-Optimized Wearable 5G Network} 
The inadequate battery volume and intensive processing requirement of the future 5G wearable networks is attracting the researchers to come up with the new energy efficiency schemes. It is obvious that the solutions based on modern energy harvesting can improve the battery life of wearable devices. For the wearable devices, the energy can be harvested in a variety of ways using the existing energy harvesting techniques.  Energy for the wearable devices can be harvested through wireless signals and through human body (body movement, body heat and body friction), to name a few \cite{8750873}. However, the higher frequency spectra and potentially multiple sources of energy-harvesting for the 5G wearable network demand the network to be energy-optimized. Further research on the selection of energy harvesting techniques and optimization of usable energy for individual categories of wearable 5G devices is required in the future. \par
\subsection{Three-tier Cooperative 5G Network with Energy Harvesting as Core Aspect}
Energy efficiency and spectral efficiency are among the most important issues in the 5G network. Most of the existing 5G architectures consider the energy and spectrum efficiency separately \cite{wu2018spectral}. A few studies also exist in the literature; which tend to integrate the energy efficiency and spectral efficiency. Furthermore, the mobility of individual 5G devices introduces ambient noise that affects both the energy efficiency and the spectral efficiency. One way of coping with this challenge is to present a three-tier cooperative 5G network while considering energy harvesting as core aspect in which the trio (a) energy efficiency, (b) spectral efficiency and c) mobility scenarios are collectively considered. The existing energy harvesting techniques can improve the energy efficiency while the efficient use of available spectra can deal with exponential growth of 5G devices. The consideration of mobility scenario will further complement the future 5G technologies by managing ambient noise. However, the cooperation among the above mentioned three tiers demand new algorithms and techniques to be developed in the future.\par
\subsection{Wireless Power Transfer}
Simultaneous Wireless Information and Power Transfer (SWIPT) essentially allows the information and power to be transmitted simultaneously using the same RF signal \cite{wu2019transceiver}. Non-orthogonal Multiple Access (NOMA) is a key enabler for 5G network that addresses various challenges such as serving multiple users over same radio resources and high throughput. Researchers have already investigated the application of SWIPT to NOMA and presented promising results in terms of increased network throughput and energy efficiency \cite{xiao2018joint}. However, the existing studies investigate the SWIPT-NOMA applications by considering idealistic assumptions such as perfect channel state information and perfect hardware. Nonetheless, 5G devices can suffer from imperfections such as hardware impairments and imperfect channel state information. However, the practical constraints such as residual hardware impairments and channel state information over multiple real-world scenarios are yet to be analyzed in detail. \par
\section{Conclusion}
\label{sec:conclusion}
Tremendous advances in wireless technologies have raised the concern of substantial energy consumption. In the foreseeable future, improving the energy efficiency of battery-equipped smart devices in 5G networks will become one of the main issues. This study was conducted in the context of utilizing energy harvesting technology to prolong the lifetime of devices and networks. In this study, we taxonomized the literature on the basis of several important parameters. The key requirements to enable energy harvesting in 5G networks for providing guidelines to the new researchers are discussed. Furthermore, some open challenges and future research directions are presented. Finally, we conclude that energy harvesting plays an important role in prolonging the battery life of devices and networks by harvesting energy from environmental sources and ambient RF signals. Thus, serious attention must be given in addressing the discussed challenges in the future.\par

\section*{Acknowledgment}
Imran's work is supported by the Deanship of Scientific Research, King Saud University through Research Group Project number RG-1435-051.

\bibliographystyle{IEEEtran}
\bibliography{references}

\begin{IEEEbiography}[{\includegraphics[width=1in,height=1.25in,clip,keepaspectratio]{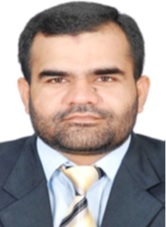}}]{Muhammad Imran} is an associate professor at King Saud University. His research interest includes MANET, WSNs, WBANs, M2M/IoT, SDN, Security and privacy. He has published a number of research papers in refereed international conferences and journals. He served as a Co-Editor in Chief for EAI Transactions and Associate/Guest editor for IEEE (Access, Communications, Wireless Communications Magazine), Future Generation Computer Systems, Computer Networks, Sensors, IJDSN, JIT, WCMC, AHSWN, IET WSS, IJAACS and IJITEE.
\end{IEEEbiography}
\begin{IEEEbiography}[{\includegraphics[width=1in,height=1.25in,clip,keepaspectratio]{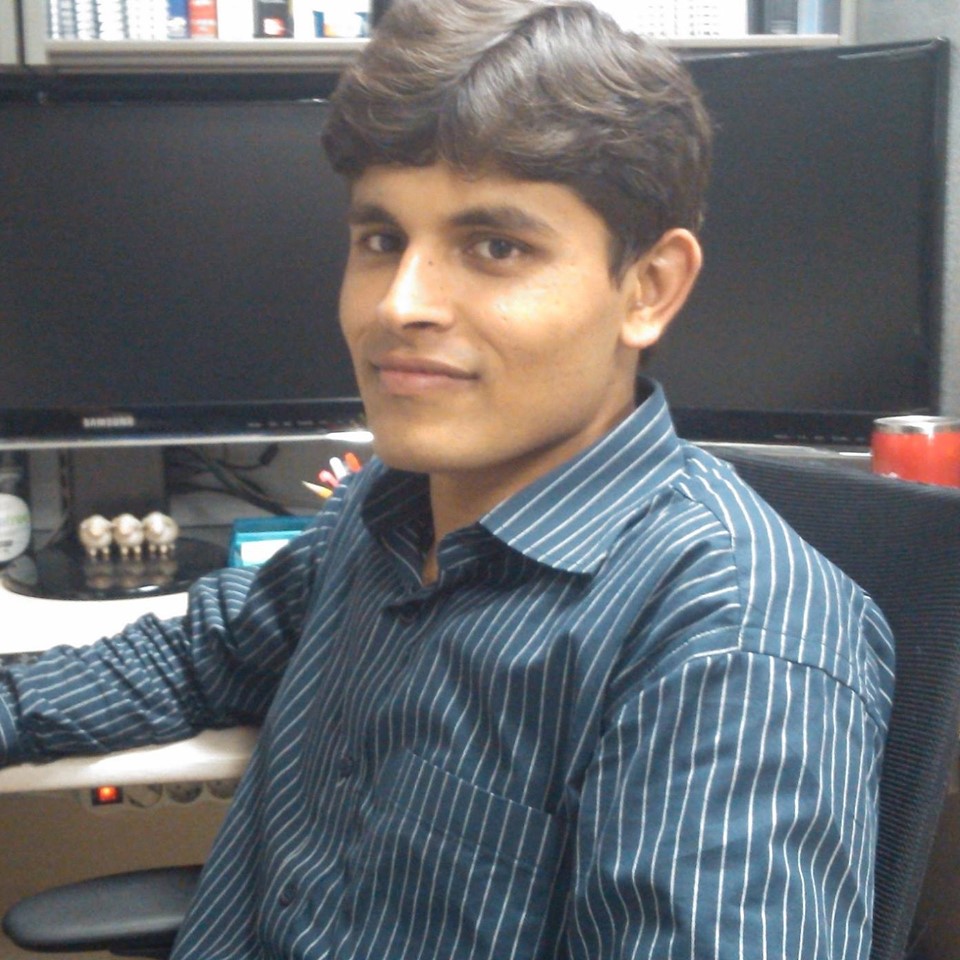}}]{Latif U. Khan} is currently working as a faculty member at Department of Telecommunication Engineering, University of Engineering \& Technology, Mardan, Pakistan. He received his MS (Electrical Engineering) degree with distinction from University of Engineering and Technology, Peshawar, Pakistan in 2017. His research interests include analytical techniques of optimization and game theory to edge computing and end-to-end network slicing.
\end{IEEEbiography}
\begin{IEEEbiography}[{\includegraphics[width=1in,height=1.25in,clip,keepaspectratio]{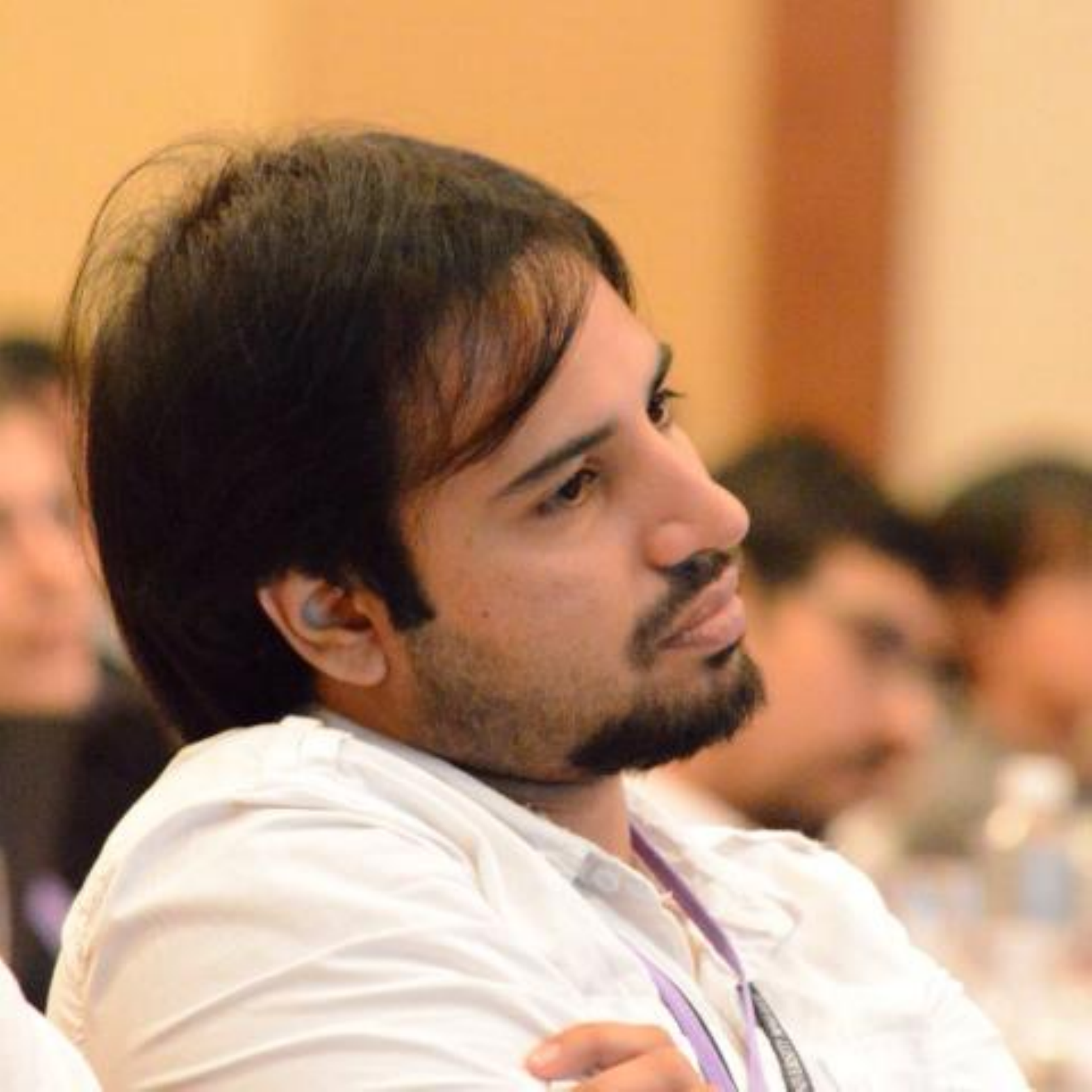}}]{Ibrar Yaqoob} (S'16, M'18, SM'19) is a research professor at the Department of Computer Science and Engineering, Kyung Hee University, South Korea, where he completed his postdoctoral fellowship. He received his Ph.D. (Computer Science) from the University of Malaya, Malaysia. He is a guest/associate editor in various Journals. He has been involved in a number of conferences and workshops in numerous capacities. His research interests include big data, mobile cloud computing, the Internet of Things, and computer networks. 
\end{IEEEbiography}
\begin{IEEEbiography}[{\includegraphics[width=1in,height=1.25in,clip,keepaspectratio]{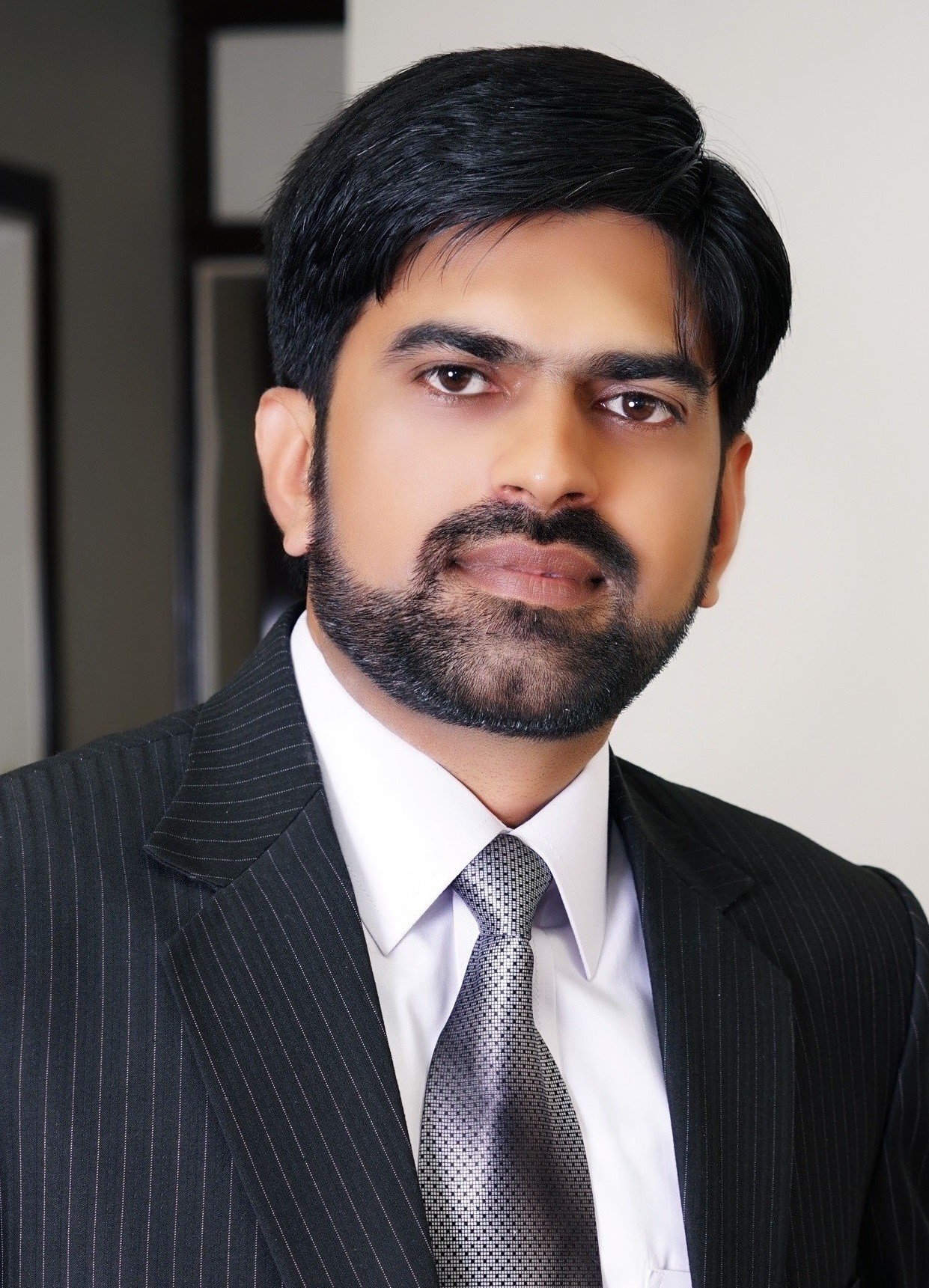}}]{Ejaz Ahmed} (S’12, M’17, SM’18) received his PhD in Computer Science from University of Malaya, Malaysia. He is Associate Technical Editor/Editor of IEEE Communications Surveys \& Tutorials, IEEE Communications Magazine, IEEE Access, Elsevier JNCA, KSII TIIS, and Elsevier FGCS. He has served as Chair and Co-chair in several international conferences. His areas of research interest include Mobile Cloud Computing, Mobile Edge Computing, Internet of Things, Cognitive Radio Networks, Big Data, and Internet of Things.
\end{IEEEbiography}
\begin{IEEEbiography}[{\includegraphics[width=1in,height=1.25in,clip,keepaspectratio]{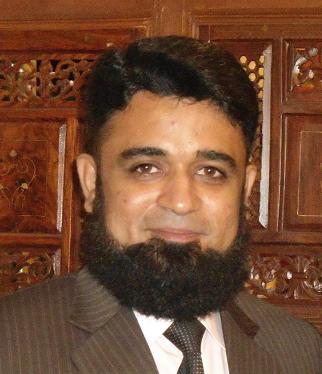}}]{Muhammad Ahsan Qureshi} received his PhD from University of Malaya, Kuala Lumpur, Malaysia in 2016. His major work is on radio propagation modeling for Vehicular Ad Hoc Networks (VANETS). He is currently serving as Assistant Professor in the Faculty of Computing and Information Technology, University of Jeddah, Khulais, Saudi Arabia. His research interests include Wireless Communication, Traffic Management, Green Computing and Internet of Vehicles.
\end{IEEEbiography}   
\begin{IEEEbiography}[{\includegraphics[width=1in,height=1.25in,clip,keepaspectratio]{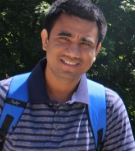}}]{Arif Ahmed} received his M.Tech. degree in computer science and engineering from the National Institute of Technology Silchar, India, in 2014. He worked as a visiting scientist at the Centre for Development of Advanced Computing, Mumbai, India, from 2014 to 2015. His research interests are in the field of mobile cloud computing, fog computing, software-defined networking, and mathematical modeling.
\end{IEEEbiography}
\balance
\end{document}